# Variable stars in the field of the open cluster NGC 7789

K. Jahn[1], J. Kaluzny[1], and S. M. Ruciński[2]

[1] Warsaw University Observatory, Al. Ujazdowskie 4, 00-478 Warsaw, Poland
[2] Institute for Space and Terrestrial Science, 4850 Keele St., Toronto, Ontario M3J 3K1, Canada

**Abstract.** Fifteen variable stars have been discovered during four nights of monitoring of NGC 7789. It was possible to determine periods for ten of these variables. Seven objects are W UMa-type binaries, in this number five are members of the cluster. Among the remaining variables, there are three further close binaries, one $\delta$ Scuti which is a Blue Straggler in the cluster, and one multi-modal variable (with indeterminable periods). One of the binaries showed an eclipse-like event; this binary is a good candidate for determination of masses of its components at the turn-off point of the cluster.

**Key words:** binaries: close – binaries: eclipsing – $\delta$ Scu – stars: variables: other – open clusters and associations: individual NGC 7789

## 1. Introduction

This paper is a contribution to the on-going systematic search for short-period variables in open and globular clusters which has been started in December 1989 and is based on time resolved CCD photometry. The main goals of this observing program were described by Kaluzny et al. (1993). Partial results of the survey in the form of results for several stellar clusters have already been published (Be 39: Kaluzny et al. 1993, Be 33: Mazur et al. 1993, NGC 4372: Kaluzny & Krzemiński 1993, NGC 6791: Kaluzny & Ruciński 1993a). A large fraction of variables discovered in those clusters are the W UMa type binaries. A compilation of the new data on contact binaries in open clusters was presented by Kaluzny & Ruciński (1993b) and Ruciński & Kaluzny (1994). These review papers discussed also possibilities of studying the evolution of binary systems and formation of Blue Stragglers on the basis of surveys like this one, and summarized frequencies of occurrence of such binaries in clusters of different age and richness.

In the present paper, we report the results for the open cluster NGC 7789 ($l = 116°$, $b = -5°$), which is much younger than most of the clusters in which significant numbers of contact binaries have been found to date. Extensive photometry of this rich, intermediate-age cluster was presented first by Burbridge & Sandage (1958), who determined colors and magnitudes of nearly seven hundred stars. The apparent distance modulus for the cluster was estimated as $(m - M)_V = 12.0$, the reddening $E(B - V) = 0.24$, the metallicity [Fe/H] = $-0.26 \pm 0.06$. The cluster age is about 2 Gyr (see Friel & Janes, 1993). We redetermined the distance modulus of the cluster. Our estimate, based on the lower limit of the red-giant clump on the color-magnitude diagram, assumed to be $M_V = 0.8$ (Castellani et al. 1992, cf. also Sect. 5 and Fig. 4), yields $(m - M)_V = 12.3$. This estimate is used in the paper.

Sections 2-4 describe briefly methods of the data reduction, of the search for variables and of period determination for discovered variables. In Sect. 5, we present the color-magnitude diagram for the cluster, and in Sect. 6, we discuss the cluster membership of the variables. The eclipsing binaries are discussed in Sect. 7 and the remaining variables in the field of NGC 7789 are presented in Sect. 8.

## 2. Observations and data reduction

The data were collected at the Kitt Peak National Observatory using the 0.9-m telescope with a 1024×1024 Tektronics CCD having a field of view of 11.6×11.6 arc min$^2$ (scale 0.68 arc sec/pixel). The cluster was monitored for about 4 hours during each of four consecutive nights in 1991 October 8–11 (UT). Part of the data was collected through clouds with seeing typically between 1″.2 and 2″.2. The cluster centre was positioned at the centre of the field of view. Most of the observing time was spent on sequential 10-minutes exposures interspersed with 50-seconds observations, both taken through the $V$ filter. We obtained 60 long and 63 short exposures in this way. Three additional frames were collected through the B filter: one 50 s and two 10 min exposures. The preliminary processing of the raw data was made within IRAF[1]. Stellar photometry was carried

*Send offprint requests to*: K. Jahn

---

[1] IRAF is distributed by the National Optical Astronomy Observatories, which are operated by the Association of Universities for Research in Astronomy, Inc., under cooperative agreement with the National Science Foundation

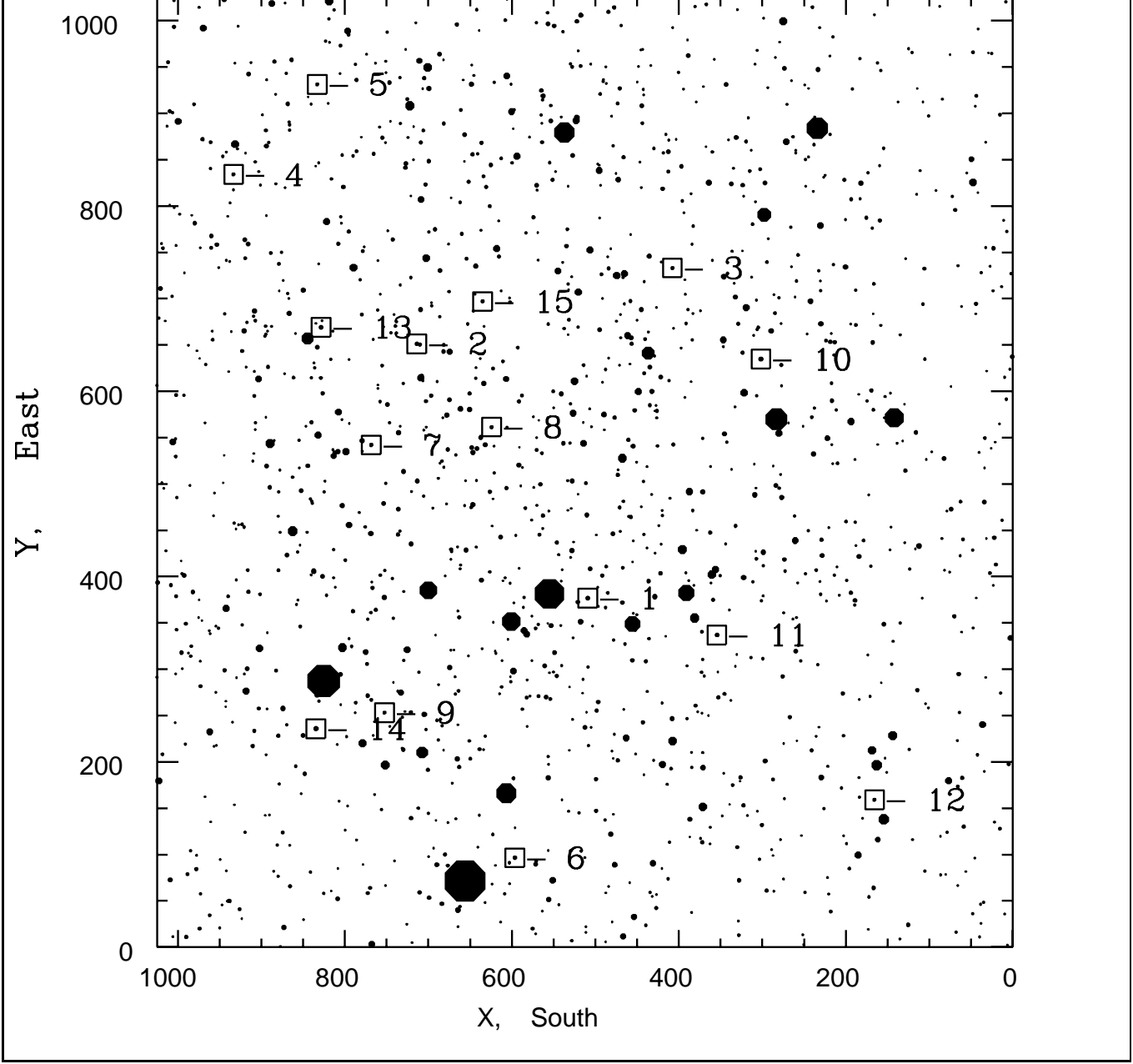

**Fig. 1.** Finding chart for the variables discovered in the field of NGC 7789. Besides the variables, only stars with $V < 18^{m}_{.}0$ have been plotted

out using DoPHOT (Version V1.1, Schechter et al. 1993). An analytical, position-independent, point spread function (PSF) was used for extraction of profile-fitted magnitudes.

## 3. Search for variable stars

The present analysis of the photometric data follows that of Kaluzny & Ruciński (1993a). The two sets of frames in the $V$ filter, i.e. long- and short-time exposures were analyzed separately in a similar manner. First, stars with trustworthy photometry have been selected on each frame. These were the objects classified as Type 1 in DoPHOT, i.e. unblended stars with relatively high signal-to-noise ratios, used in computing the analytical model of the PSF. The number of such objects on 10 min frames ranged from about 3350 on the best frames to about 2600 on the poorest ones, while on the 50 s frames these numbers were 2150 and 1930, respectively. These variations in the photometric quality were caused by passing clouds.

Coordinates of stars measured on individual frames were next transformed to a common system. Subsequently, we constructed a master list containing all detected stars, and then determined a light curve for each star on this list. The detailed description of the method of selection of variables and of the extraction of light curves was given by Kaluzny et al. (1993). A star was selected as a candidate variable if its light curve showed a dispersion significantly larger than for other stars of similar magnitude. Moreover, it had to be measured on at least 45 frames for each set of exposures (10 min or 50 s). 48 stars satisfying these conditions were selected for further analysis.

**Table 1.** Rectangular and equatorial coordinates (based on GSC, Lasker et al. 1988) of variables in the field of NGC 7789. $X$ and $Y$ coordinates correspond to positions on the finding charts (see Fig. 1)

| Id | X | Y | $\alpha$(2000) (h:m:s) | $\delta$(2000) (deg:m:s) |
|---|---|---|---|---|
| 1 | 508.95 | 376.94 | 23:57:09.95 | 56:42:17.7 |
| 2 | 713.67 | 651.34 | 23:57:33.51 | 56:44:32.6 |
| 3 | 407.60 | 733.20 | 23:57:39.49 | 56:40:59.7 |
| 4 | 933.16 | 834.22 | 23:57:49.45 | 56:46:59.6 |
| 5 | 832.80 | 931.25 | 23:57:57.28 | 56:45:48.0 |
| 6 | 596.22 | 96.61 | 23:56:46.67 | 56:43:23.7 |
| 7 | 768.17 | 542.08 | 23:57:24.50 | 56:45:12.7 |
| 8 | 624.01 | 561.50 | 23:57:25.73 | 56:43:32.8 |
| 9 | 752.24 | 253.16 | 23:57:00.23 | 56:45:08.1 |
| 10 | 301.57 | 634.84 | 23:57:30.98 | 56:39:49.0 |
| 11 | 354.25 | 337.13 | 23:57:06.21 | 56:40:31.9 |
| 12 | 165.53 | 159.20 | 23:56:50.81 | 56:38:25.7 |
| 13 | 828.36 | 669.17 | 23:57:35.33 | 56:45:51.3 |
| 14 | 834.51 | 235.93 | 23:56:59.00 | 56:46:05.2 |
| 15 | 634.80 | 696.98 | 23:57:37.10 | 56:43:37.1 |

**Table 2.** Parameters of variables in the field of NGC 7789. $V_{\max}$ is an estimated magnitude from the observed maximum light, $A_V$ is the range of observed variations in the $V$ band. The internal errors of colors $< B - V >$ range from $0\overset{m}{.}01$ for the brightest stars to $0\overset{m}{.}04$ for the faintest ones. The period is given in days.

| Id | Type | $V_{\max}$ | $A_V$ | $< B - V >$ | Period |
|---|---|---|---|---|---|
| 1 | EW | 14.26 | 0.16 | 0.68 | 1.19 |
| 2 | EW | 14.81 | 0.16 | 0.69 | 0.72 |
| 3 | EW | 15.30 | 0.07 | 0.69 | 0.70 |
| 4 | EW | 16.72 | 0.14 | 1.05 | 0.337 |
| 5 | EW | 19.32 | 0.40 | 1.20 | 0.2387 |
| 6 | EW | 14.54 | 0.05: | 0.62 | 0.884 |
| 7 | EW/EB | 15.93 | 0.06 | 0.93 | 0.455 |
| 8 | EB | 15.19 | 0.22 | 0.65 | 0.85 |
| 9 | EA | 19.10 | >0.35 | 1.06 | 0.776 |
| 10 | $\delta$ Scu | 14.04 | >0.03 | 0.56 | 0.0955 |
| 11 | ? | 14.45 | >0.12 | 0.70 | ? |
| 12 | ? | 16.13 | >0.08 | 1.23 | ? |
| 13 | ? | 14.35 | >0.05 | 0.69 | ? |
| 14 | ? | 14.01 | 0.04 | 2.06 | ? |
| 15 | ? | 15.47 | 0.06: | 0.54 | ? |

After detailed examination of all candidates, we rejected spurious variables and a few possible faint variables with very noisy light curves. Finally, we selected 15 certainly variable stars. Their rectangular and equatorial coordinates are listed in Table 1. The transformations to the equatorial system are based on the Guide Star Catalogue (Lasker et al. 1988), with errors less than $0\overset{\prime\prime}{.}4$ in $\alpha$ and less than $0\overset{\prime\prime}{.}2$ in $\delta$. Figure 1 shows the finding chart, which simplifies the identification of all variables. The $X$ and $Y$ coordinates on this chart correspond to the coordinates in Table 1.

### 4. Periods and light curves

Periodicities of all discovered variables were analyzed with a program based on the CLEAN algorithm (Roberts et al. 1987). About half of our variables exhibit continuous, quasi-sinusoidal variations which are typical for the W UMa-type stars. For such light curves, which have two minima of similar depth, the algorithms based on Fourier techniques tend to find the maximum power at frequencies twice those of the real ones. We allowed for this effect and then refined the period determinations using the analysis of variance method, as described by Schwarzenberg-Czerny (1989). This method was used also for other variables.

We were able to determine the periods of 10 variables: seven W UMa-type systems, two binaries of the type EB and EA, and one $\delta$ Scuti type star. Some of period determinations are preliminary and need improvement, especially those for V6 and V10. The periods we adopted are listed in Table 2, which also gives the type of variability, the maximum brightness observed in the $V$ band, the amplitude of variations, and the $< B - V >$ colors. Phased light curves of stars V1-V10 are shown in Fig. 2. More detailed discussion of all eclipsing binaries discovered in the field of NGC 7789 is presented in Sect. 7.

Periods and types of the remaining five variables cannot be identified with confidence. We show their $V$ magnitude as a function of time in Fig. 3. One of these stars, V11, is most probably an EA-type eclipsing binary as its light curve exhibited a descending-branch feature and was flat outside this presumed eclipse event. Another one, V12, seems to have a period only slightly shorter than 1 day so that our observations covered a very small fraction of its orbit. We find also a multi-modal oscillating star (V15), but do not have enough data to determine its period(s).

The other two variables, V13 & V14, showed mean brightness levels changing from night to night. Although the magnitude of these changes did not exceed $0\overset{m}{.}07$, they seemed to be real judging by the high quality of the data for these stars. We discuss this group of variables in more detail in Sect. 8.

Tables containing full photometry for all stars discussed in this work are available in electronic form from the Centre de Donness de Strasburg (CDS); see the editorial in A&A 1993, Vol. 280, page E1.

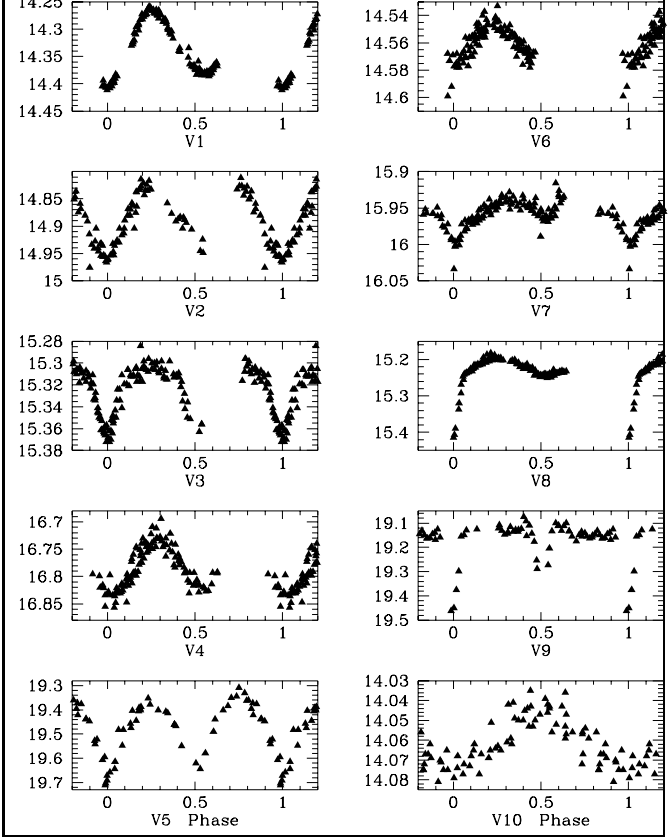

**Fig. 2.** Phased $V$ lights curves of eclipsing binaries (V1-V9) and $\delta$ Scuti-type variables found in the field of NGC 7789

## 5. The CMD and colors of the variables

We used our best-quality frames to to construct the $V$ versus $B$–$V$ color-magnitude diagram (CMD) for the observed field of NGC 7789. Thus, we used two $B$ frames, three long-exposure, and three short-exposure $V$ frames. We created a master list of all stars measured on these frames. The instrumental magnitudes were calculated as weighted averages of magnitudes measured on individual frames.

Observations of eight standard stars from the field around Ru 149 (Landolt 1992) and 14 stars from the field near M 92 (Stetson & Harris 1988) were used to determine the color terms of the transformation from the instrumental to the standard $BV$ system. The transformation adopted has the following form:

$$v = V + 0.046 \times (B - V) + \text{const},$$
$$b - v = 0.843 \times (B - V) - 0.03 \times (B - V) \times X + \text{const},$$

where the lower-case letters refer to the instrumental magnitudes. The zero points of the photometry were determined using additional photometry obtained with the KPNO 2.1-m telescope (Kaluzny, in preparation). The colors of the variables were determined by interpolating the $V$ magnitudes to the moments of mid-exposures of the $B$ frames (as each $B$ exposure was bracketed by two $V$ exposures). We assumed that colors as representative for our variables with an uncertainty of about $0^m_.05$, which is the maximum range of variations of $B - V$ in typical W UMa systems. This uncertainty should not be larger for the detached binaries, since none of the above exposures falls near the brightness minimum, i.e. near the possible substantial change of colors due to eclipses.

The $V$ magnitudes at the maximum brightness and the average $< B - V >$ indices for the variables in NGC 7789 are listed in Table 2 together with the amplitudes of variability (or their lower limits). Figure 4 shows the CMD for the cluster, with the positions of variables marked at their maximum $V$ brightness using their average color indices. The squares represent eclipsing binaries (EW, EB and EA), the open circle marks the $\delta$ Scuti-type variable, and the asterisks mark the variables of unknown type.

## 6. Cluster membership of the variables

Four of the variables have proper motions and the membership probabilities determined by Mc Namara & Solomon (1981). The probabilities are listed in the column "PMP" in Table 3. According to these data, the W UMa-type system V6 is a field star but this does not agree with the $M_V(\log P, B - V)$ calibration (next section) which predicts the brightness of the star $M_V = 2^m_.05$. This is only $0^m_.21$ above the observed value for the assumed $(m - M)_V = 12.3$, and indicates that the star is at the distance of the cluster. Could it be that the system is now leaving the cluster and hence has a large proper motion? Or, conversely, could it have been captured by the cluster from the Milky Way field?

The $M_V(\log P, B - V)$ calibration's predictive power does not exceed about $\pm 0^m_.5$, which translates into about $\pm 25\%$ spread in distances. Thus, V6 might be in front or behind the cluster by that much. This clearly shows that the $M_V$ calibration is useful for testing null hypothesis (i.e. as a sieve to remove non-members) but is not as useful for assigning memberships. We should remember however, that since the cluster is in the Milky Way plane, with high density of stars, there is a finite probability that the star is indeed close to or even crossing the cluster, yet does not belong to it.

The remaining three objects, the $\delta$ Scuti star V10 and the unclassified objects V11, and V13 are members of the cluster with the membership probabilities higher than 90%.

## 7. Eclipsing variables

We discovered seven W UMa systems in the cluster field. They show quite typical EW-type light curves except for V3 which shows minima somewhat narrower than normally seen in EW variables. Possibly, it is a close but detached system (EA). In turn V7 shows unequal minima and we classify its light curve as EW/EB. We have applied the absolute brightness calibration established recently by Ruciński (1994) to the seven W UMa-type variables. The calibration, utilizing simple linear dependencies on $\log P$ and $B - V$, is able to eliminate field interlopers when deviations $M_V^{obs} - M_V^{cal}$ exceed about one magnitude. Among our contact systems only one object, V5, seems to be such an interloper. As we commented already, to our surprise V6 does agree with the calibration to $0\overset{m}{.}2$ although its proper motion is discordant with other members of the cluster. For the (possibly) detached system V3, the calibration predicts $M_V$ by $0\overset{m}{.}4$ brighter than observed which might be an indication that the system is indeed underfilling its Roche lobes.

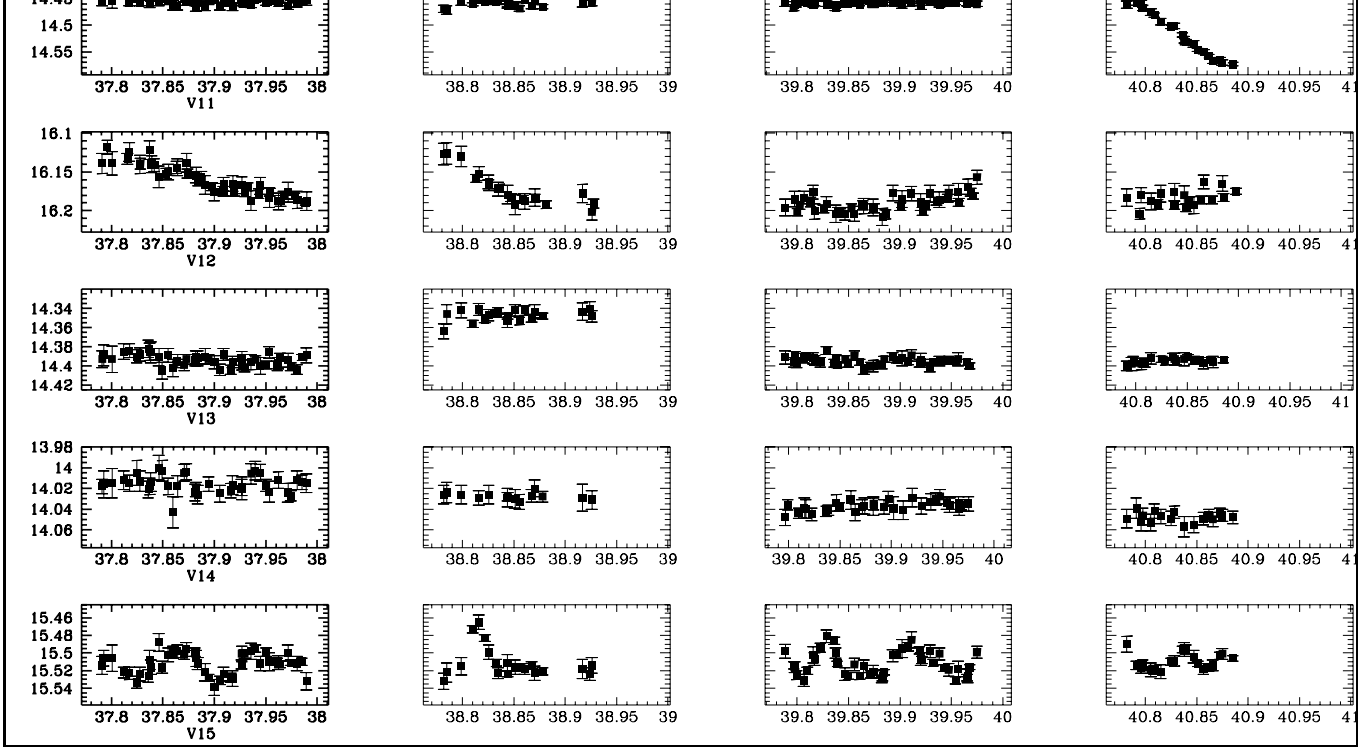

**Fig. 3.** Light curves of variables V11-V15 for which the variability types cannot be established. Time is expressed in HJD – 2448500

**Table 3.** Membership of discovered variables: probability PMP according to proper motions (Mc Namara & Solomon, 1981), and the absolute brightness calibration for the W UMa-type systems (Ruciński, 1994)

| Id | PMP | $M_V^{obs} - M_V^{cal}$ |
|----|------|-------|
| 1  | –    | –0.03 |
| 2  | –    | –0.03 |
| 3  | –    | +0.43 |
| 4  | –    | –0.44 |
| 5  | –    | +1.17 |
| 6  | 0.00 | +0.21 |
| 7  | –    | –0.40 |
| 10 | 0.95 | –     |
| 11 | 0.90 | –     |
| 13 | 0.94 | –     |

We note that none of the five W UMa-type systems members of NGC 7789 can be considered as a blue straggler but two contact binaries V1 and V2 have colors and magnitudes that place them near the turn-off on the cluster CMD, with absolute magnitudes of about $M_V \simeq 2\overset{m}{.}0 - 2\overset{m}{.}5$.

Positions of the contact binaries V4 and V7 on the CMD and their periods indicate that these objects have properties similar to evolved W UMa-type systems such as AH Vir, OO Aql or those discovered in NGC 188 by Kaluzny & Shara (1987). In particular the light curve of V7 is very similar to V5 in NGC 188 which can be considered as a good prototype of evolved contact systems.

We should note that none of the W UMa-type systems except the non-member V5, exhibits variations larger than $0\overset{m}{.}16$. Most probably, as interpreted by Kaluzny & Ruciński (1993a), the CCD searches of open clusters offer a better statistics of variability amplitudes than the sample of bright contact binaries in the sky field which is incomplete and biased towards larger amplitudes. However, at this moment, we cannot exclude a tantalizing possibility that contact systems in clusters have different origins and systematically smaller mass-ratios than the field systems, which would result in smaller amplitudes of light variations.

The lower limit to the frequency occurence of W UMa systems can be estimated by simply dividing the number of discoveries by the number of the cluster members. We analyzed

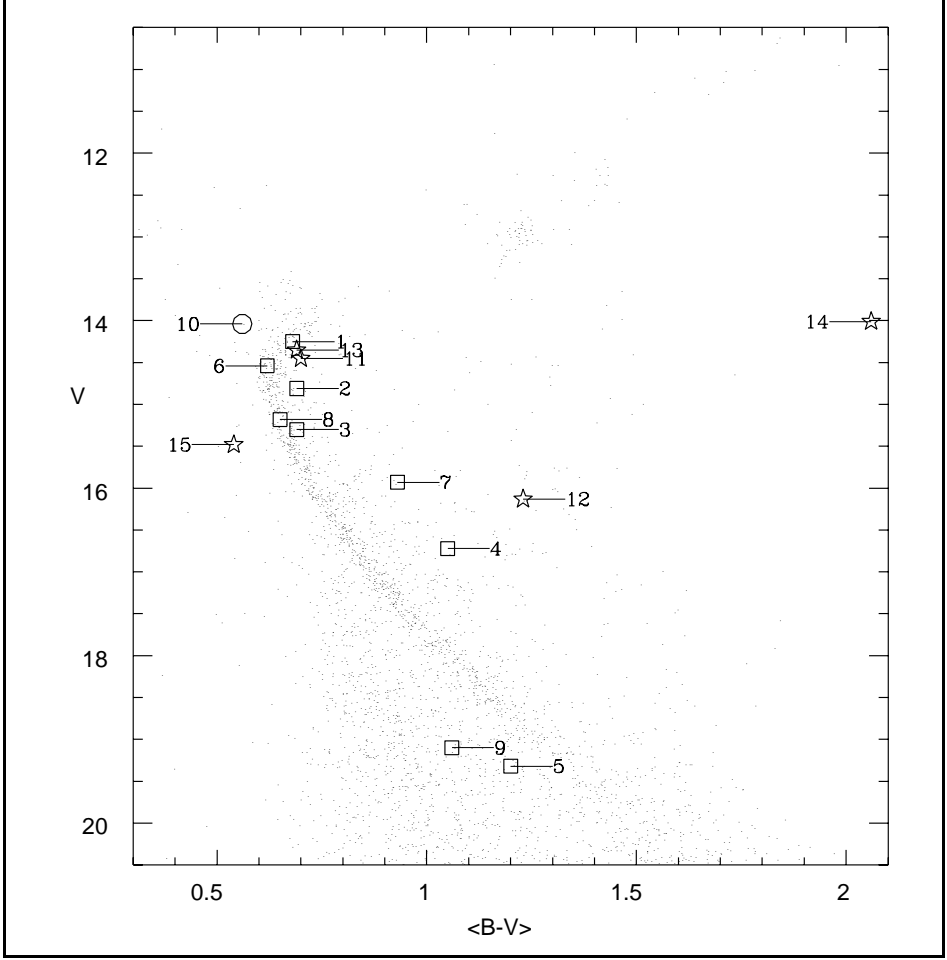

**Fig. 4.** The color-magnitude diagram of NGC 7789 with the positions of the variables (labelled by their Id numbers). Squares correspond to eclipsing binaries, the open circle marks a $\delta$ Scuti-type variable, and asterisks represent variables of unknown type

about 900 stars with $V < 19\rlap{.}^m 0$ in the cluster and discovered five (or six, depending on the inclusion of V6) W UMa systems among them. Assuming a uniform probability of detection as a function of magnitude and of amplitude we can estimate the lower limit for the frequency of occurence of contact binaries in the cluster as 0.0056 (or 0.0067). Note, that the good quality of our light curves for $V < 17\rlap{.}^m 0$ makes it rather unlikely that somewhat fainter binaries (down to $V < 19\rlap{.}^m 0$) would escape the detection.

Now, turning to the remaining eclipsing binaries in the cluster. The variable V9 lies probably behind the cluster, as suggested by its position on the CMD – about $1\rlap{.}^m 0$ below the main sequence. This is an interesting object for it exhibits modulations of brightness (with a period of 1.5–2 hours, and amplitude $\approx 0\rlap{.}^m 04$) superimposed on its light curve outside the well defined minima. The relatively short orbital period ($0\rlap{.}^d 77$) seems to leave little room for oscillations of one of the components. Perhaps it is a cataclysmic variable, but then of an unusually red color.

We were not able to determine the periods of V11 and V12. The variable V11 belongs to the cluster with the probability 90% according to the proper motion criterion. The V magnitude and the color of V11 places this object on the turn-off of the cluster. The light curve shows a distinct eclipse event, but our observations do not cover the whole period of this binary. Further monitoring of this object should allow to determine its full light curve. It would be also worthwhile to determine the radial velocity curves for this binary. Then, the masses of the components of the binary at the turn-off for the cluster could be derived.

The phase coverage of the light curve of V12 is unfortunate, since it has probably a period of about one day, and it happened that we measured its brightness near its brightness minima only. Should it be a member of the cluster, it would be unlikely that it is a close binary with evolved components, like the variables V4 or V7, since the observed period would be then too long.

## 8. Remaining variables

The $\delta$ Scuti variable V10 belongs to the cluster with the probability 95%. The absolute magnitude $1\rlap{.}^m 64$ and the color $0\rlap{.}^m 32$ place this object in the Dwarf Cepheid instability strip (e.g. Mateo 1993), so that V10 is a pulsating blue straggler in the cluster. It is worthwhile to note that until recently only two such objects were found in 'old' open clusters with ages greater than 2–3 Gyr (see the review of Mateo, 1993).

Another variable (V15), which is probably also a $\delta$ Scuti type star most probably is not a member of the cluster, as suggested by its position on the CMD. Its light curve shows variations with a period of about 2 hours (see Fig. 4), but there is also some modulation which makes impossible to determine the variability periods from our short time-base data.

The variable V13 is a cluster member with the probability of 94% (see Table 2). It lies near the turn off of the cluster. We observed only a brightening of this object during the second night (see Fig. 3), so we were not able to identify characteristics of its variability.

V14 showed a systematic decrease of brightness during the four consecutive nights. Thus, its period is longer than 4 days. The membership of this object cannot be established, but its position on the CMD might suggest that it is not a member of the cluster. Probably, this is a late type spotted star in the field of NGC 7789, for its $<B-V>$ is very large.

## 9. Conclusions

A systematic search for variable stars in the moderate-age cluster NGC 7789 has led to discovery of 15 variable stars in the cluster field. Five W UMa systems are members of the cluster, but none is located in the blue straggler area of the CMD. However, most of the eclipsing variables that we discovered are located near the turn-off region which confirms the expected high probability for discovery for binaries with evolving components. One of the eclipsing systems showed brightness changes characteristic for a total eclipse; as a member of the cluster, this system would be an excellent candidate for an in-depth photometric and spectroscopic study.

*Acknowledgements*. KJ and JK were supported by the KBN grant 2 1177 91 01. Research of SMR was supported by a grant from the Natural Sciences and Engineering Council of Canada. Partial support from the ESO C&EE grant No. A-01-065 is also acknowledged with gratitude.